\documentclass[doublecol]{epl2} 
\title{Robust oscillations in SIS epidemics on adaptive networks:
Coarse-graining by automated moment closure}
\shorttitle{Coarse Graining and Oscillations in an adptive SIS model} 
\author{Thilo Gross\inst{1}\inst{2} \and Ioannis G.~Kevrekidis\inst{3}}
\shortauthor{T.~Gross and I.G.~Kevrekidis}

\institute{                    
\inst{1} Dept.~of Chemical Engineering, Princeton University, Princeton NJ 08544, USA.\\
\inst{2} Max-Planck Institute f\"ur Physik komplexer Systeme, N\"othnitzer Stra\ss e 38, 01187 Dresden, Germany. email: gross@pks.mpg.de\\
\inst{3} Dept.~of Chemical Engineering and PACM, Princeton University, Princeton NJ 08544, USA.
  email: yannis@princeton.edu\\
}

\abstract{
We investigate the dynamics of an epidemiological susceptible-infected-susceptible (SIS)
model on an adaptive network. This model combines epidemic spreading (dynamics {\em on} the network) 
with rewiring of network connections (topological evolution {\em of} the network).
We propose and implement a computational approach that enables us to study the 
dynamics of the network directly on an emergent, coarse-grained level.
The approach sidesteps the derivation of closed low-dimensional approximations.
Our investigations reveal that global coupling, which enters through the awareness of the population to 
the disease, can result in robust large-amplitude oscillations of the state and topology of the 
network.  
}

\begin{document}

\maketitle

\section{Introduction}
%% INTRODUCTION %%%%%%%%%%%%%%%%%%%%%%%%%%%%%%%%%%%%%%%%%%%%%%%%%%%%%%%%%%%%%%%%%%%%%%%%%%%%%%%%%%%%%%%%%%%%%%%%%%%%%%
Over recent years the physics of networks has received much attention \cite{Dorogovtsev:Review,Barabasi:Review,Newman:NetworkBible}. 
Research has focused on the one hand on dynamics {\em of} networks, i.e., network growth and rewiring \cite{Barabasi:Scaling,Watts:SmallWorlds}, and on the other hand on dynamics {\em on} networks; the coupling of individual dynamical systems according to a static network topology \cite{PastorSantorras:ScaleFree,May:ScaleFree}. 
Only recently have these two distinct strands of network research been brought together in the study of \emph{adaptive networks}, which combine topological evolution of the network with dynamics on the network nodes \cite{Bornholdt:AdaptiveBoolean, Zhou:AdaptiveNets, Ito:Leaders, Traulsen:Mapping, Newman:Social, epidemics,anreview}. 
Adaptive networks have been shown to exhibit a number of new phenomena, including robust self-organization towards dynamical criticality\cite{Bornholdt:AdaptiveBoolean}, the formation of complex global topologies based on simple local rules \cite{Sneppen:SelfAssembly,epidemics}, the emergence of new bifurcations and phase transitions involving local as well as topological degrees of freedom \cite{epidemics,Newman:Social}, and finally a spontaneous `division of labor' in which an initially homogeneous population of nodes self-organizes into functionally distinct classes\cite{Ito:Leaders}. 

A prominent application of networks is epidemiology. In an ever more populated and tighter connected world 
epidemic diseases are once again becoming a major threat \cite{Karlen:Epidemics}. In the past decades we have witnessed the return of old diseases \cite{Oldstone:History} as well as the emergence of new ones \cite{Omi:Sars} at an accelerating pace. 
Conceptual mathematical models can provide fundamental insights on the underlying mechanisms that govern epidemic dynamics. 

One of the most simple conceptual models of epidemics is the susceptible-infected-susceptible (SIS) model \cite{Anderson:Epidemics}, described in more detail below. 
If considered on a static network, varying the parameters in this model reveals at most one dynamical transition. 
This transition is continuous and corresponds to the epidemic threshold: the point in parameter space beyond which the disease can invade the network. 
If one enables the individuals in the model to avoid contact with infected ones by altering their local interaction topology the system turns into an adaptive network. 
It has recently been shown that this natural extension of the spatial SIS model gives rise to more complex  dynamics, including bistability, hysteresis and a narrow region of oscillatory dynamics \cite{epidemics}. 

In order to understand the subtle interplay between local dynamics and topological evolution that characterizes adaptive networks the application of tools from dynamical systems theory is desirable. 
While these methods are in general only applicable to relatively low-dimensional {\em equation-based} models, previous results \cite{epidemics,Ehrhardt:AdaptiveSocial} suggest that often a small number of topological degrees of freedom suffices to characterize the dynamics of the individual-based model on an emergent level. 
This means that a small number of observables can describe the macroscopic (system-level) state of the network and thus a low-dimensional description at this level is feasible.
In \cite{epidemics} a low-dimensional set of system-level equations of motion for the adaptive SIS model was derived by means of a moment expansion and subsequent moment closure approximation. 
While this approach has yielded satisfactory results, it involves a strong homogeneity assumption and therefore can fail in certain topologies. 

%% WHAT WE DO %%%%%%%%%%%%%%%%%%%%%%%%%%%%%%%%%%%%%%%%%%%%%%%%%%%%%%%%%%%%%%%%%%%%%%%%%%%%%%%%%%%%%%%%%%%%%%%%%%%%%%%%%
In this Letter we extend the previous work in two ways. 
First, we consider the effect of a variable level of awareness to the disease. This awareness is shared instantaneously among the population and therefore acts as a global coupling. 
Second, we propose and implement an alternative computational approach, based on the coarse-grained `equation-free' modeling and analysis framework \cite{Kevrekidis:Review}.
In this approach the analytical moment-closure approximation is replaced by a numerical procedure which automatically extracts the effects of appropriate closure.
This approach enables us to investigate the emergent dynamics of complex networks with well-established tools of dynamical systems theory {\em without} explicitly deriving a closed analytical description on this level.
Our investigation reveals that the global coupling strongly extends the parameter region in which oscillatory dynamics can be observed. 
It thereby gives rise to persistent large-amplitude oscillations affecting the prevalence of the disease as well as the topology of the network.  

%% THE MODELL %%%%%%%%%%%%%%%%%%%%%%%%%%%%%%%%%%%%%%%%%%%%%%%%%%%%%%%%%%%%%%%%%%%%%%%%%%%%%%%%%%%%%%%%%%%%%%%%%%%%%%
\section{The adaptive SIS model}
We consider a network with a fixed number of nodes $N$ and undirected links $L$. 
Every node represents an individual, while links represent social contacts between individuals.
An individual can be either infected (I) or susceptible (S). 
We denote the links between individuals as SS-links, SI-links or II-links according to the states of the nodes that they connect. 
In every time step infected individuals recover with a probability $r$, becoming susceptible. 
For every SI-Link there is a probability $p$ that the susceptible individual becomes infected. 
In addition to these standard rules of SIS models, we allow susceptible individuals to avoid contact with infected individuals by altering their local interaction topology:
For every SI-Link there is a probability $w$ that a rewiring event occurs in a given time step. 
In a rewiring event, the susceptible individual cuts its link to the infected individual and establishes a new link to another susceptible individual.
In contrast to \cite{epidemics} we do not fix the rate $w$, but assume that it changes according to the individual's awareness of the disease. 
In human populations with access to mass media it is reasonable to assume that information on the disease is instantaneously disseminated across the network. We therefore set $w=w_0 \rho$, where $\rho=i/N$, is the infected fraction of the population and $w_0$ is a constant parameter. 

\section{Coarse graining}
%% ANALYTICAL MOTIVATION, BACKGROUND AND DEFINITIONS %%%%%%%%%%%%%%%%%%%%%%%%%%%%%%%%%%%%%%%%%%%%%%%%%%%%%%%%%%%%%%%%%%
It has been shown in \cite{epidemics} that the system-level dynamics of sufficiently large networks (here $N = 10^5, L=10^6$) can be captured by the number of SS-links $l_{\rm SS}$, the number of II-links $l_{\rm II}$ and the number of infected individuals $i$. 
The number of susceptibles $s$ and the number of SI-links $l_{\rm SI}$ then follow from the conservation laws $s+i=N$ and $l_{\rm SS}+l_{\rm SI}+l_{\rm II}=L$.
A moment expansion reveals that the dynamics of the collective state variables $s, l_{\rm SS}, l_{\rm II}$ can be described by 
\begin{equation}  
\begin{array}{r c l}
\frac{\rm d}{\rm dt}i&=& p l_{\rm SI} - r i \label{eqMoments}\\ 
\frac{\rm d}{\rm dt}l_{\rm II}&=& p l_{\rm SI}\left(\frac{l_{\rm ISI}}{l_{\rm SI}}+1\right)-r i\frac{2l_{\rm II}}{i} \\
\frac{\rm d}{\rm dt}l_{\rm SS}&=& r i \frac{l_{\rm SI}}{i} - p l_{\rm SI} \frac{l_{\rm SSI}}{l_{\rm SI}}+w l_{\rm SI} 
\end{array}
\end{equation}
where $l_{\rm ISI}$ and $l_{\rm SSI}$ denote the number of I-S-I and S-S-I chains in the network, respectively.

Note that we have so far only assumed that the network is large, so that the state can be described by continuous variables and the effect of individual-based (demographic) stochasticity can be neglected.
In particular no strong homogeneity has been used since many effects of non-homogeneous topology    
are captured by taking the densities of links $l_{\rm SS}, l_{\rm SI}, l_{\rm II}$ and chains $l_{\rm ISI}, l_{\rm SSI}$ into account. However, Eq.~(\ref{eqMoments}) does not yet constitute a closed model, as we haven't specified how $l_{\rm ISI}$ and $l_{\rm SSI}$ are determined.

In Ref.~\cite{epidemics} the moment-closure approximation $l_{\rm ISI}=l_{\rm SI}^2/s, l_{\rm SSI}=l_{\rm SS}l_{\rm SI}/2s$ was used to close the model. 
This approximation introduces a homogeneity assumption as it requires that 
the SI- and SS-links are homogeneously distributed among the possible three node chains. 
While the moment closure approximation has yielded good results in \cite{epidemics} it has to fail 
in networks with slowly decaying degree distribution, such as scale free graphs.

%% COARSE ANALYSIS: BASIC PROCEDURE %%%%%%%%%%%%%%%%%%%%%%%%%%%%%%%%%%%%%%%%%%%%%%%%%%%%%%%%%%%%%%%%%%%%%
In this Letter we avoid the moment closure approximation:
We do \emph{not} attempt to derive analytical expressions for $l_{\rm ISI}$ and $l_{\rm SSI}$ in order to close Eq.~(\ref{eqMoments}).
Instead, we specify a numerical procedure that generates the effect of appropriate closure terms \emph{on-demand} from short bursts of numerical simulation. 
Therefore the major technical challenge in this Letter is to determine the correct value of $l_{\rm ISI}$ and $l_{\rm SSI}$ for a given set of state variables $i, l_{\rm SS}, l_{\rm II}$.

In general one can image a large number of \emph{candidate} network topologies that agree with the desired values of $i, l_{\rm SS}, l_{\rm II}$. Thus, for a given set of $i, l_{\rm SS}, l_{\rm II}$, many different values of $l_{\rm ISI}$ and $l_{\rm SSI}$ are in principle possible. 
However, most of the candidate topologies will not arise in the long term dynamics of the network.
In fact one finds that the $l_{\rm ISI}$ and $l_{\rm SSI}$ that are consistent with given $i, l_{\rm SS}, l_{\rm II}$ are restricted to a small set \emph{in the long-term dynamics}. 
      
Consider the following: Large adaptive networks can easily have millions of topological degrees of freedom.
The assumption that the state of such a network can be characterized by a low number of variables implies in general that there is a time scale separation between a few slow variables and the remaining fast degrees of freedom.  
The system will then approach a slow manifold, on which the long term dynamics takes place. The slow manifold can be parameterized entirely by the slow variables, as all other topological degrees of freedom are slaved to the slow variables.
Thus a set of slow variables determines the state of the system. 
This implies that all other degrees of freedom can be written as a function of the slow variables.   

In the system considered here the time scale separation between the slow variables $i, l_{\rm SS}, l_{\rm II}$ and fast variables, such as $l_{\rm SSI}$ and $l_{\rm ISI}$, is finite and thus the reasoning given above is only approximately true. 
Nevertheless, the values of $l_{\rm ISI}$ and $l_{\rm SSI}$ that can possibly appear in the long term dynamics (while a given set of $i, l_{\rm SS}, l_{\rm II}$ is observed) are restricted to a relatively narrow band.
This band is simply a noisy graph of the slow manifold.   
If the time scale separation is sufficiently large, the band of possible $l_{\rm ISI}$ or $l_{\rm SSI}$ is narrow enough to treat it effectively as a single value. 
In the following we denote such a value of $l_{\rm ISI}$ or $l_{\rm SSI}$ that is consistent with a given set of $i, l_{\rm SS}, l_{\rm II}$ in the long term dynamics simply as a \emph{consistent value}.  

\begin{figure}
\centering
\includegraphics[width=3in,height=1.8in]{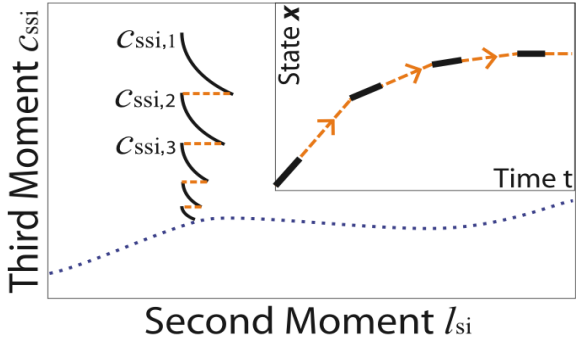}
\caption{(Color online) Schematic of the lifting procedure. A network with initially arbitrary
third moments is settled to the slow manifold (dotted line) on which the consistent values encountered 
in the long-term dynamics are located. For this purpose we run a short simulation (solid line) and then 
reset the first and second moments while retaining the third moments (dashed line). Inset: Illustration of 
coarse projective integration. The necessary information for long projective leaps (dashed) is extracted from 
short bursts of individual-based simulation (solid).
\label{figSchema}}
\end{figure}

Let us now describe an algorithm that computes the consistent values of $l_{\rm ISI}$ and $l_{\rm SSI}$ for a given set of state variable values $i^*$, $l_{\rm SS}^*$, $l_{\rm II}^*$. 
In order to account for the combinatorial effect of the slow variables, it is advantageous to work with the 
normalized third moments $c_{\rm ISI}=sl_{\rm ISI}/{l_{\rm SI}}^2$, $c_{\rm SSI}=l_{\rm SSI}s/(l_{\rm SS}l_{\rm SI})$. 
We start by randomly generating a network at the desired operating point $i^*$, $l_{\rm II}^*$, $l_{\rm SS}^*$. The third moments $l_{\rm ISI}$ and $l_{\rm SSI}$ in a random graph will not, in general, be consistent, that is, they are not on the slow manifold of the system.
In order to ``settle" the network to the slow manifold, where the consistent values are located, we proceed as follows: 
We run a short simulation of $j$ steps. 
Over the course of the simulation the system approaches the (attracting) slow manifold; the values of 
the third moments at the end of the simulation provide us with a first estimate $c_{{\rm ISI},1}$, $c_{{\rm SSI},1}$.
To obtain a better estimate, we generate a network at the operating point $i^*$, $l_{\rm II}^*$, $l_{\rm SS}^*$; with third moments according to $c_{{\rm ISI},1}$, $c_{{\rm SSI},1}$ (explained below). 
The endpoint of a second burst of simulation yields a better estimate $c_{{\rm ISI},2}$, $c_{{\rm SSI},2}$. 
This process is repeated $n_{\rm init}$ times to converge on the slow manifold (see Fig.~\ref{figSchema}).
This is followed by $n_{\rm aver}=n_{\rm tot}-n_{\rm init}$ further repetitions.
Averaging over the endpoints of these repetitions we compute the consistent values $c_{\rm x}^*=\sum_{n=n_{\rm init}}^{n_{\rm tot}} c_{{\rm x},n}/n_{\rm aver}$, for x=ISI, SSI. 
The variance of the approximation and therefore the width of the band in which the consistent values are located can be estimated by $\sum_{n=n_{\rm init}}^{n_{\rm tot}} ({c_{\rm x}^*} - {c_{{\rm x},n}})^2/({n_{\rm aver}}^2-n_{\rm aver})$.   

An important step in the algorithm outlined above is the generation of a network with specific $i,l_{\rm SI},l_{\rm SS},l_{\rm SSI}$, and $l_{\rm ISI}$. Starting from a given network the desired number of $i,l_{\rm SI} $ and $l_{\rm II}$ can be reached straightforwardly by first flipping the state of nodes and then selectively rewiring links. Then we set $l_{\rm SSI}$, $l_{\rm ISI}$: We chose a random link and additionally a random node that is not part of the link. 
If at least one of the nodes connected to the link has the same state as the individual node we selected, we can 
rewire the link in a way that it leaves $i,l_{\rm SI}$ and  $l_{\rm II}$ unchanged 
but can possibly affect $l_{\rm SSI}$ and $l_{\rm ISI}$. If this rewiring brings us closer to 
the desired values it is accepted, otherwise a new node and a new link is chosen randomly. This process is repeated until the desired $l_{\rm SSI}$ and $l_{\rm ISI}$ have been reached.        

\begin{figure}
\centering
\includegraphics[width=3in]{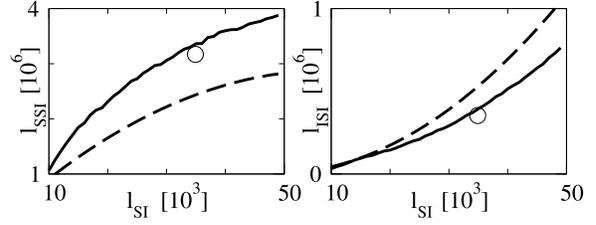}
\caption{ A slice of the slow manifold ($i=97693$, $l_{\rm ss}=884707$). 
Shown are the closure terms $l_{\rm ssi}$ and $l_{\rm isi}$
computed by the proposed procedure (solid line) and an analytical approximation (dashed line).
The corresponding results from a long simulation run (circles) agree well with the numerical result.
$N=10^5$, $L=10^6$, $w_0=0.006$, $p=0.0007$, $r=0.0002$, . 
\label{figManifold}}
\end{figure}

\section{Numerical Methods}
The algorithm described above allows us to generate consistent states for Eq.~(\ref{eqMoments}) computationally on-demand. 
This enables us to use Eq.~(\ref{eqMoments}) to numerically estimate the temporal derivatives of the system-level state variables at a given point in state space.
It thereby provides us with all the information that is needed to apply many established tools for the computational investigation of dynamical systems.   

Let us consider a simple example. A standard task in the investigation of dynamical systems is the computation 
of a trajectory, starting from a given initial point ${\bf x}_0$. 
Perhaps, the most simple computational tool to perform this task is forward Euler integration: Given a point ${\bf x}_n$ on the trajectory a subsequent point ${\bf x}_{n+1}$ is computed by the first order Taylor approximation ${\bf x}_{n+1}={\bf x}_n+\tau \left. f({\bf x})\right|_{{\bf x}_n}$, where
$f({\bf x})$ is the right-hand side of the equations of motion and $\tau$ is a fixed increment in time. 
By repeated application of this procedure a sequence of points is produced that traces the desired trajectory.     
The approach of coarse-grained projective integration \cite{Kevrekidis:Overview} can be used to apply methods like forward Euler integration to systems in which no closed analytic expression for the right-hand side of the equations of motion is available.
Following this approach the derivatives, such as $\left.{\rm d}{\bf x}/{\rm dt}\right|_{\bf x_n}=\left.f({\bf x})\right|_{\bf x_n}$, are not computed analytically, but are estimated from properly initialized bursts of microscopic simulation. 

We have performed coarse projective integration of the SIS model considered here. We have also used a variant scheme that takes some additional information from the moment expansion into account:     
In order to compute the temporal derivative of the state variables at a given point in state space ${\bf x}=(i,l_{\rm II},l_{\rm SS})$, we first use the procedure described in the previous section to compute appropriate $l_{\rm ISI}$ and $l_{\rm SSI}$ for Eq.~(\ref{eqMoments}). 
We can then close Eq.~(\ref{eqMoments}) and use it to estimate the temporal derivatives.  
Thus, in every step of this projective Euler method {\em a new network is generated} in order to compute the derivatives of the system-level variables. 
However, in order to construct the network and find the desired closure terms only short bursts of simulation are necessary; no long simulation runs are required. 

\begin{figure}[htb]
\centering
\includegraphics[width=2.5in,height=2in]{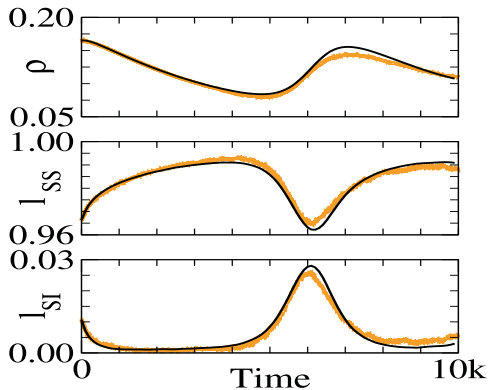}
\caption{(Color Online) Coarse projective Integration. Shown are time series of the infected fraction of the 
population, the density of SS-links $l_{\rm SS}$ and SI-links $l_{\rm SI}$. The link numbers have been normalized with respect to the total number of links in the network. The results from a coarse projective integration (black) are in good agreement with a full individual-based simulation (orange). Parameters: $N=10^5$, $L=10^6$, $p=0.06$, $r=0.0002$, $p=0.0003$, $n_{\rm  init}=2$, $n_{\rm aver}=1$, $s=20$, $\tau=100$.  
\label{figProjective}}
\end{figure}

A comparison of the coarse projective integration scheme with full individual-based simulation is shown in Fig.~\ref{figProjective}.
Coarse projective integration has previously been proposed as a way to speed up lattice Boltzmann, kinetic Monte Carlo as well as molecular dynamics simulations \cite{Kevrekidis:Overview}. 
In the present implementation for networks, initialization of consistent networks is time consuming, and thus only a small speed-up is realized (a factor of $\sim2$).
However, a significantly bigger speedup can be expected in larger networks. 
In addition there are other advantages: 
In the derivation of Eq.~(\ref{eqMoments}) the individual-based based stochasticity of the system is to a large extend filtered out. 
Furthermore, the coarse integration scheme can run backwards in time, which can for instance be useful in approximating the boundaries of basins of attraction \cite{Kevrekidis:Reverse}.
But most importantly the projective Euler integration is a proof of principle; many more powerful methods of dynamical systems theory can be applied in essentially the same way.             

%% NEWTON AND BIFURCATIONS %%%%%%%%%%%%%%%%%%%%%%%%%%%%%%%%%%%%%%%%%%%%%%%%%%%%%%%%%%%%%%%%%%%%%%%%%%%% 

The simple example of forward Euler integration proves that it is possible to close the model on-demand by using only short bursts of microscopic simulation. 
This allows us for instance to use fixed-point algorithms such as Newton's method. 
These methods do not only compute stationary states more efficiently than integration to stationarity; 
they can also be used to find dynamically unstable steady states, such as saddles, that are inaccessible to direct simulation. 

Starting from an initial estimate ${\bf x}_0$ of a stationary state, Newton's method produces
progressively better estimates by ${\bf x}_{n+1}={\bf x}_n-{{\rm \bf J}}^{-1}f({\bf x}_n)$, where 
${\rm \bf J}$ is the Jacobian matrix at ${\bf x}_n$. 
As the convergence of our Newton method was not very sensitive to minor errors in the Jacobian
we found it advantageous to estimate $f({\bf x})$ by the automated closure procedure and approximate the Jacobian by differentiating the (inaccurate but faster) analytical closure.

The Jacobian is also needed for continuation of a solution branch and for detection of bifurcations
\cite{Kuznetsov:Elements}. 
Since the eigenvalues can be highly sensitive to errors in the Jacobian we approximate the elements of Jacobian in this case by finite differences of temporal derivatives computed numerically. 
Because of the sensitivity of the eigenvalues, averaging over many ($\sim 10^3$) temporal derivative evaluations is necessary to obtain reasonably accurate eigenvalues. 
A possibly better alternative is the application of matrix-free iterative algorithms, in which the eigenvalues are determined without explicit approximation of the Jacobian \cite{Kelley:Newton}.

\section{Results}
%% RESULTS %%%%%%%%%%%%%%%%%%%%%%%%%%%%%%%%%%%%%%%%%%%%%%%%%%%%%%%%%%%%%%%%%%%%%%%%%%%%%%%%%%%%%%%%%%%%%

\begin{figure}
\centering
\includegraphics[width=3in]{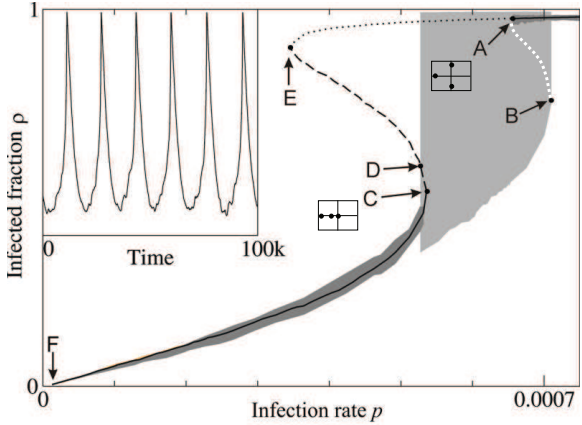}
\caption{System-level bifurcation diagram. Branches of stable steady states (solid lines) and 
saddle points (dashed/dotted) have been computed by a coarse-grained Newton method. The white line 
tentatively marks a branch of saddle limit cycles. 
Shaded regions mark ranges of $\rho$ observed during 
long individual-based simulations in the neighborhood
of the attractive limit cycle (light gray) and of stable
of stationary solutions (dark gray).
Computation of the Jacobian eigenvalues reveal a 
subcritical Hopf bifurcation (A), two fold 
bifurcations (C,E) and a transcritical bifurcation (F). In addition there is a fold bifurcation of cycles (B) 
and a homoclinic bifurcation (D). Two small insets indicate the eigenvalue configuration at points A and C.
Inset: time series on the limit cycle attractor at $p=0.0006$.
Parameters: $w_0=0.06$, $r=0.0002$, $N=10^5$, $L=10^6$. 
\label{figBif}}
\end{figure}

By applying the Newton method for several parameter values, we have computed a one-parameter bifurcation diagram for the adaptive SIS network (Fig.~\ref{figBif}).
For $p>0.71$ the network approaches a stationary state in which the prevalence of the diseases is high $\rho\approx 0.97$. 
If $p$ is lowered, a threshold is crossed (point B) at which a coarse limit cycle attractor emerges -- most likely from a discontinuous non-equilibrium phase transition which appears as a fold bifurcation of limit cycles in the low dimensional system.
We expect that a coarse saddle cycle is also born there. 
The stationary attractor and the limit cycle attractor coexist until the stability of the stationary attractor is lost in a subcritical Hopf bifurcation (point A). 
In such bifurcations an unstable limit cycle vanishes. 
While we were not able to compute this unstable oscillation directly, it is consistent with the saddle cycle mentioned above.

At lower infection rates a stable steady state and a saddle point emerge from a fold bifurcation (point C).
This discontinuous transition marks the edge of a narrow interval in which the limit cycle attractor coexists with the new stationary attractor.
At slightly lower infection rates the limit cycle vanishes in what appears as a homoclinic bifurcation (point D). 
After this bifurcation the steady state that has emerged from C is the only attractor of the system. 
The two saddles vanish in a fold bifurcation at point E.
Finally, at the epidemic threshold (point F), the stable equilibrium undergoes a transcritical bifurcation with the trivial equilibrium ($i=l_{\rm II}=l_{\rm IS}=0$) as it leaves the positive cone of the state space. 

Let us emphasize that Fig.~\ref{figBif} describes the dynamics \emph{on the emergent level}: in the non-trivial stationary states the network is {\em never frozen}, but constantly undergoes changes in its topology and the states of its nodes. 
However, these changes are balanced in such a way that system-level variables, such as $i$, $l_{\rm SS}$ and $l_{\rm II}$ remain effectively stationary. 
Furthermore, note that two of the state variables ($l_{\rm SS}$ and $l_{\rm II}$) describe topological degrees of freedom. Without adaptive rewiring, it is well known that the emergent level dynamics of the SIS model can be described by the single variable $i$ and only exhibits transition E \cite{Anderson:Epidemics}.
Dynamics such as the self-organized cycles for SIS model are therefore only possible on adaptive networks.     

The dynamics of the adaptive SIS model with variable awareness reported here differ from the dynamics of the model with constant rewiring studied in \cite{epidemics}. Rewiring is in both cases a strong protective mechanism that isolates infected individuals and thus reduces the prevalence of the disease. 
However, in the present model this mechanism only plays the role if the prevalence and therefore also the population's awareness of the disease is high. 
In particular, compared to the non-adaptive SIS model, rewiring does not increase the epidemic threshold (point F) in the present model, while a strong increase of the threshold was found in the model with constant rewiring. 
Perhaps the most striking difference between the model variants is that the awareness mechanism greatly increases the parameter range in which oscillations can be observed. 
Also, the amplitude of the oscillations is increased considerably.

In order to understand how variable awareness promotes oscillations, the underlying mechanism has to be taken into account. 
While rewiring of network connections isolates infected individuals and thus slowly decreases the prevalence of the disease, it also leads to the formation of a highly connected cluster of susceptible individuals (see \cite{epidemics} for details). 
If the disease manages to invade this cluster it can rapidly propagate causing a large outbreak. This leads back to a state of high prevalence completing the cycle.     
In the model with constant rewiring the invasion of the susceptible cluster is difficult as it is still protected by rapid rewiring. 
If the amplitude of the cycle is large the number of infected individuals at the low point is small, which reduces the chance of successful invasion further. 
Therefore large amplitude oscillations cannot persist in the model with constant rewiring and the parameter range in which oscillations are observed is small. 
In the model studied in this Letter large oscillations also result in small numbers of infected individuals in the low point of the cycle. 
However in this case the disease is unlikely to become extinct as the system is still beyond the epidemic threshold and the protection offered by rewiring decreases proportional to the number of infected. 
Since the disease free state is thus dynamically unstable reinfection of the susceptible cluster is bound to occur.

%% CONCLUSIONS %%%%%%%%%%%%%%%%%%%%%%%%%%%%%%%%%%%%%%%%%%%%%%%%%%%%%%%%%%%%%%%%%%%%%%%%%%%%%%%%%%%%%%%%
\section{Summary and Conclusions}
In this Letter we have proposed a computational approach to the investigation of emergent properties of adaptive networks. 
By applying the methods of coarse-grained, `equation-free' modeling and analysis we have shown that existing numerical tools of dynamical systems theory can be used to investigate networks directly on an emergent level. 
The approach enabled us to obtain information that is not directly accessible by simulation alone, while avoiding the often prohibitively difficult derivation of closed system-level equations of motion. 
In particular we have avoided the strong homogeneity assumption that is inherent in previous analytical moment closure approximations.

Our approach depends critically on the assumption that \emph{bona fide} emergent-level variables exist and can be identified; else closure (numerical or otherwise) at the desired level is impossible.
The moment-based description used here is applicable to networks in which the number of different states of the nodes is small and the topological change can be expressed in terms of the network moments.
The success of this description depends critically on the time scale separation between the state variables and the higher moments of the network.  
In our model the time scale separation is finite and decreases as the rewiring rate is increased. 
At very high rewiring rates the computation of closure terms ceases to converge.

In other systems it may be advantageous to use different sets of system-level variables, such as heuristic variables, based on the researcher's experience or variables identified by automated data-mining algorithms \cite{Kevrekidis:DiffusionMaps}. 
Although dynamical equations, such as Eq.~(\ref{eqMoments}) are in this case not available, one can still estimate temporal derivatives of the chosen set of variables directly from short bursts of properly initialized simulation. 

Here we only applied two simple numerical tools: Forward Euler integration and the Newton method.
These tools revealed significant information on the coarse bifurcation structure of the system. 
Demonstrating the applicability of such tools to a network problem constitutes a proof of principle. 
More sophisticated integrators and fixed point algorithms, continuation of solution branches, automated bifurcation detection, computation of normal form coefficients etc. can {\em in principle} be analogously applied. 
These tools are available in the form of free, well-tested software packages that are commonly used for the investigation of systems of ordinary differential equations. 
By writing numerical wrappers that operate along the lines discussed above, this arsenal of existing, highly efficient methods can be brought to bear on the investigation of adaptive networks. 

In the present Letter we have used the proposed procedure to investigate an epidemiological SIS model.  
In our model individuals can protect themselves by altering their topological neighborhood at a rate determined by their awareness to the disease.
Our analysis shows that in a considerable parameter range the prevalence of the disease and the topology of the network exhibits oscillations of large amplitude.
Let us emphasize that the observed oscillations are not \emph{caused} by the variable level of awareness, since in our model awareness changes without time lag. 
Instead variable awareness levels extend the parameter range in which oscillations can be observed; decreasing awareness levels at low prevalence stabilize cycles which would otherwise drive the disease to extinction. 
The simple mechanism studied in our conceptual model may explain oscillatory dynamics observed in certain real world diseases.

\acknowledgments
This research is partially supported by the Humboldt Foundation (TG), DARPA and DTRA (IGK, TG).

\bibliographystyle{eplbib}
\bibliography{nld,OnNetworks,OfNetworks,NetworkReview,AdaptiveNetworks,MyPapers,Coarse,Numerics,epidemics}
\end{document}